\documentclass[10pt,conference]{IEEEtran}
\IEEEoverridecommandlockouts

\usepackage{cite}
\usepackage[T1]{fontenc}
\usepackage{amsmath,amssymb,amsfonts}
\usepackage{algorithmic}
\usepackage{graphicx}
\usepackage{textcomp}
\usepackage{xcolor}
\usepackage{url}
\def\BibTeX{{\rm B\kern-.05em{\sc i\kern-.025em b}\kern-.08em
    T\kern-.1667em\lower.7ex\hbox{E}\kern-.125emX}}
\usepackage{flushend}

\begin{document}

\title{Treat societally impactful scientific insights as open-source software artifacts}

\author{\IEEEauthorblockN{Cynthia C.~S.\ Liem}
\IEEEauthorblockA{\textit{Multimedia Computing Group} \\
\textit{Delft University of Technology}\\
Delft, The Netherlands \\
c.c.s.liem@tudelft.nl \\
https://orcid.org/0000-0002-5385-7695}
\and
\IEEEauthorblockN{Andrew M.\ Demetriou}
\IEEEauthorblockA{\textit{Multimedia Computing Group} \\
\textit{Delft University of Technology}\\
Delft, The Netherlands\\
a.m.demetriou@tudelft.nl \\
https://orcid.org/0000-0002-0724-2278}
}

\maketitle

\begin{abstract}
So far, the relationship between open science and software engineering expertise has largely focused on the open release of software engineering research insights and reproducible artifacts, in the form of open-access papers, open data, and open-source tools and libraries. In this position paper, we draw attention to another perspective: scientific insight itself is a complex and collaborative artifact under continuous development and in need of continuous quality assurance, and as such, has many parallels to software artifacts. Considering current calls for more open, collaborative and reproducible science; increasing demands for public accountability on matters of scientific integrity and credibility; methodological challenges coming with transdisciplinary science; political and communication tensions when scientific insight on societally relevant topics is to be translated to policy; and struggles to incentivize and reward academics who truly want to move into these directions beyond traditional publishing habits and cultures, we make the parallels between the emerging open science requirements and concepts already well-known in (open-source) software engineering research more explicit. We argue that the societal impact of software engineering expertise can reach far beyond the software engineering research community, and call upon the community members to pro-actively help driving the necessary systems and cultural changes towards more open and accountable research.
\end{abstract}


\section{Introduction}
This article is a `paper'\footnote{Most likely, it will not reach the reader on paper, but as a digital PDF.}. At the moment it will reach broader readership with a formal citation attached, it will have passed peer review, and be part of a referenceable collection of proceedings of the ICSE 2023 Software Engineering in Society Track. This form and workflow have been the traditional template for communicating scientific outcomes, where getting papers accepted at prestigious venues has traditionally been treated as the major indicator of academic achievement.

Academic research has been operating under scarcity, both regarding job and research funding security. As a consequence, (not) getting major publications accepted and sufficiently cited thus has great career consequences. Still, for a long time, research communities have been acknowledging that contributions of scientific insight extend much beyond a paper, and proposals for open science have emerged, including ventures into open access, open and FAIR (Findable, Accessibile, Interoperable, Reusable) data, and open-source software.

The software engineering research community has been acting upon this~\cite{Mendez2020}, with open science policies now being explicit parts of well-respected venues like ICSE and the Empirical Software Engineering Journal, open-source tools with artifact badging being explicitly encouraged, and the option to submit registered reports entering several sub-communities such as the Conference on Mining Software Repositories. Software engineering researchers also have actively contributed to discussions on applying FAIR principles to research software~\cite{fairsoftware}.

With this position paper, we wish to inspire the software community to look even beyond this. More specifically, considering empirical scientific insights in the broad sense (i.e., insights requiring empirical observation of phenomena, often expressed in the form of data measurements), we will argue that making these insights more open will require infrastructure and quality assurance mechanisms similar to those needed in developing complex open-source software artifacts.

\section{Arguments for open science beyond the paper}
Already in 1942, Robert K.\ Merton noted that anti-intellectualism was rising and the integrity of science was under attack. In response, four `institutional imperatives' were formulated as comprising the ethos of modern science: \emph{universalism} (the acceptance or rejection of claims entering the lists of science should not depend on personal or social attributes of the person bringing in these claims), \emph{``communism''} [sic] (common ownership of scientific findings, with the imperative to communicate findings, as opposed to secrecy), \emph{disinterestedness} (upholding scientific integrity by not having self-interested motivations), and \emph{organized skepticism} (judgment on the scientific contribution should be suspended until detached scrutiny is performed, according to institutionally accepted criteria)~\cite{merton1942science}. Many scientists still subscribe to these norms today~\cite{mertioniannormsandersonetal}. These imperatives also implicitly echo in today's calls, manifestos and proposals for open science and open access~\cite{manifestoreproducible, openaccesstennant}, which push for better science, which more people can access---but with which more people also can actively interact. Below, we further elaborate on several arguments and initiatives that argue that open science should not stop at a paper that more people can read.

\subsection{Insufficient quality control on papers}
Open-access publishing may stimulate academic and societal uptake, transform the business models of publishers, and allow for publicly funded knowledge to be publicly available. Still, open access is only an aspect of open science, and insights and methods reported in a paper may not trivially be reproducible or replicable\footnote{Definitions of `reproducibility' and `replicability' have not always been used in crisp ways; e.g., compare the former~\cite{acmformer} and current~\cite{acmcurrent} ACM definitions, in which definitions are swapped. Generally spoken, in the current discussion, we do not need a sharp distinction, and rather want to refer to the overall concept that repeating an experiment should give consistent results.}, either because common specifications are not sufficiently detailed~\cite{mcfee_et_al_2019}, or because claims may be outright false~\cite{ioannidisfalse}. While researchers have been divided on which domains suffer from reproducibility crises~\cite{bakerreproducibility}, generally, many well-published works have failed to replicate in psychology~\cite{reproducibilitypsych} and cancer biology~\cite{reproducibilitycancerbiology}, and many concerns are arising about the replicability of machine learning outcomes~\cite{reproducedpapers, mlreproducibility}. This leads to credibility crises, in which it is unclear whether results can actually be trusted and built upon. When policy-makers seek to base decisions on scientific insights, this can have severe consequences to human health and public trust~\cite{ivermectin, vandersluijscredibility}.

Officially, science should be self-correcting; through peer review and active continuous scrutiny processes, illegitimate claims should be detected and corrected. However, in practice, self-correction turns out painfully slow and reluctant~\cite{vazireholcombeselfcorrecting,correctiontoolate}. This may have to do with `publish or perish' cultures being too strong in institutions, leading to unhealthy working environments~\cite{publishorperishrawatmeena,publishorperishderondmiller,publishorperishvandalenhenkens}, incentivizing Questionable Research Practices~\cite{NSRI_QRP}, and de-incentivizing investment in Responsible Research Practices~\cite{NSRI_RRP}.

\subsection{Joint resource investments for collaborative momentum}
With machine learning research, growing power and resource imbalances are observed between large industrial labs, and small labs in public institutions. A researcher at a university will likely not have sufficient computational resources and comprehensive data access to easily be able to replicate results as reported by big tech industry. Thus, joint investments in shared computation infrastructure are needed~\cite{hellendoornsawantDL}.

In psychological science, joint efforts have been coordinated into massive replication projects, where multiple teams tried to replicate canonically reported outcomes in parallel. Good examples of this are the five `Many Labs' large-scale replication projects~\cite{manylabs1, manylabs2, manylabs3, manylabs4, manylabs5}.

For such efforts, the joint investments need to focus on technical and intellectual infrastructure: i.e., the efforts required to reach a joint insight or paper, in such a way that many can indeed participate, without the transaction costs of getting started growing too large on an individual party. In other words, the focus needs to be on facilitating a shared process, rather than claiming limited-ownership output, which our present-day incentive systems still appear to push for.

\subsection{Challenges when crossing disciplines}
When research becomes interdisciplinary or even transdisciplinary~\cite{oecdtransdisciplinary}, methodology and consequent quality assurance mechanisms become more ambiguous than in the case of monodisciplinary work. While in the software engineering community, the SIGSOFT empirical standards~\cite{sigsoftempiricalstandards} help articulating and standardizing what a reviewer should expect for different types of methodological contributions, when multiple disciplines are represented at the same time, a discipline-specific reviewer may only be capable of doing a thorough quality assessment for the parts of the contribution within their expertise, but not of the full intellectual work.

In case of transdisciplinary work, a broader spectrum of stakeholders (that may not be academics) will be involved. This again causes ambiguity on how work should be reviewed and evaluated. At the same time, for societally relevant application domains, it has been argued that broader participation of stakeholders can help getting out of credibility crises with regard to modeling choices~\cite{vandersluijscredibility}. Furthermore, if academic insights are to be implemented in society, it is not unreasonable to not only push the view of academics, but also actively involve the perspectives and experiences of non-academic societal stakeholders who will be experiencing the impact of this implementation.

\subsection{Societal relevance causes vulnerability}
Research on urgent, societally relevant challenges (e.g., climate change, public health) tend to be situated in dynamic, complex, socio-technical contexts, and require transdisciplinary approaches~\cite{oecdtransdisciplinary}. Problems of relevance may be wicked~\cite{rittel1973} or even super wicked~\cite{levin2012}, meaning that there is ambiguity on how the problem should be framed (while the solution depends on the framing), and one can assess whether a solution is `better' or `worse', but there are no hard binary outcomes of whether a result is absolutely `true' or `false'. In case of super wicked problem, there is high urgency and time is running out, while there is a lack of central authority.

Acting under such dynamic uncertainty comes with challenges. While fast open publishing and knowledge-sharing can be further enabled through open science, too-hasty conclusions that have not been deeply reviewed may cause hazards to human safety~\cite{correctiontoolate, ivermectin}. Furthermore, while the general public will demand high accountability on societally impactful outcomes, at the same time, ambiguity, uncertainty, and dynamically changing insights make it impossible to end up with static, firm insights. Potentially contradicting readings on topics requiring deeper expertise can cause feelings of uncertainty in people, harming credibility of scientific work and leading to distrust~\cite{chang2015motivated}. Distrust in science causes vulnerability to credibility attacks. Indeed, in Big Tobacco, health, climate change, and AI, concerted delegitimizing efforts have been taking place as part of lobbying processes towards non-public interests~\cite{disinformationplaybook,antilla05,greyhoodieproject}. Here again, more public transparency on how insights were obtained may help in sustaining trust and facilitating broader public scrutiny.

\section{More holistic open science: from tools to conceptual parallels to open-source software}
In response to movements towards more open science, in recent years, a plethora of process improvements with supporting platforms and tools have emerged, that support releasing a more holistic scientific artifact than a paper alone. These include pre-registration (e.g., The Center for Open Science (COS)\footnote{\url{https://www.cos.io/}}, AsPredicted\footnote{\url{https://aspredicted.org}}), pre-print publication (e.g., arXiv\footnote{\url{https://arxiv.org/}}, COS), storage of additional materials beyond the PDF (e.g., COS, data repositories such as Zenodo\footnote{\url{https://zenodo.org/}} and ResearchHub\footnote{\url{https://www.researchhub.com/}}), the co-publication of research code or software artifacts (e.g., Papers with Code\footnote{\url{https://paperswithcode.com/}})), decomposed publication (e.g., Octopus\footnote{\url{https://www.octopus.ac}}, ResearchEquals\footnote{\url{https://www.researchequals.com}}, Desci Foundation\footnote{\url{https://descifoundation.org/}}), open peer review (e.g., F1000Research\footnote{\url{https://f1000research.com/}}) and pre-print / post publication peer review (e.g., PubPeer\footnote{\url{https://pubpeer.com}}, PREreview\footnote{\url{https://prereview.org/}}, Sciety\footnote{\url{https://sciety.org}}). Organizations like the COS and Psychological Science Accelerator\footnote{\url{https://psysciacc.org/}} have coordinated big-team data collection efforts. In parallel, traditional publication venues have started accepting more modern publication formats, such as registered reports~\cite{nosek2014registered}. This tooling space is presently fragmented, capturing different aspects that should improve openness in science. At a higher level, as discussed below, we however see clusters of intended functionality, that are very close to well-researched topics in software engineering research.

\subsection{Inclusive contributorship with credit}
As opposed to the traditional authorship model of publication (where author names in a list denote some undisclosed contribution to the work, the list of authors is final, and author order may imply local hierarchies that are specific to a research sub-community), there is a need to be more specific and transparent about collaborators' contributions to the intellectual work. In the publication world, the Contributor Roles Taxonomy (CRediT) has been proposed and increasingly adopted as a possible taxonomy for this, with an explicit change from authorship to contributorship~\cite{credit}.

Models of contributorship have naturally been implemented, facilitated and acknowledged in open-source software. In case multiple contributors work on the same artifact, version control systems (typically, Git) will be employed that help tracking the degree and provenance of changes (i.e., who contributed what at what time on the development timeline). Contributors may work in parallel, both working on main features needing priority, but also on more experimental features. Through branches, this can be done while there still is a consensus of what currently is a working non-breaking artifact on the main branch.
While parallel work may be done, version control systems have protocols for resolving potential conflicts arising from parallel contributions and changes. Regardless of the status of the branch, the history of contributions will always be transparent. In addition, they allow for `orphan' components of unfinished projects to also be gathered and transparently disclosed. In psychology, attention has for long been drawn to the `file drawer' problem~\cite{filedrawer}. Here, many studies with non-significant results may never have been reported, but still provide useful insights, and can help meta-scientific understanding of whether results reported as significant are indeed significant, or may have resulted from sampling bias.

We can see a similar parallel to the building of scientific knowledge: a main branch can represent current stable insights, where other branches may represent work in progress, that down the road can make the overall artifact better. Where in software engineering, code review practices ensure quality control whenever a change is to be committed (regardless of whether this is on a main or experimental branch), the same can hold for peer review, where elevated reviewing safeguards can be implemented for merging into the main branch and `pushing to production'. As we will discuss further down, the concept of the `main branch' and versioned releases has parallels to scientific consensus of current state-of-the-art.

Where in terms of ownership, public open-source repositories may have an active team of maintainers and owners of an artifact, other people not in these groups are explicitly welcome to raise issues or feature requests if they see points for improvement, and implement and suggest contributions themselves, that the maintainers and owners may choose to incorporate. Similarly, in scientific insight, a core team may work on a particular project, but other researchers and interested parties may suggest changes or improvements that could be incorporated with visible provenance.

Where open-source projects that actively seek public contributors will have clear documentation and guidelines on how to get started and contribute if one is an outsider, similar inclusion-facilitating practices can transfer to scientific research projects, as already have been demonstrated in e.g.\ the Many Labs large-scale replication projects.

\subsection{Decomposition into maintainable units}
As discussed, potential reviewers to scientific work may not naturally be equipped to thoroughly review every aspect of a complete paper, especially if this paper reflects interdisciplinary work. Generally spoken, it seems unnatural to only review a complex intellectual contribution only at release time. With pre-registration and registered reports, publishing cultures already tried to solicit such feedback earlier, with positive effects on research quality and integrity~\cite{registeredreportquality}; however, this still involves the review of complete experimental setups.

In the software engineering world, it has generally been seen as an example of good practice to organize a complex software artifact into smaller, clearly scoped modules and functions. When committing code contributions to this overall artifact, commits also would be organized in smaller, logical contributions with a clear focus, and code review would iteratively be solicited on these small contributions. This reviewing model resembles the tools facilitating decomposed publication. In software engineering, we have already seen that decomposition will help in fostering maintainability of the overall artifact, and making it easier on new contributors to quickly get onboarded on the parts of the artifact where they wish to contribute.

We explicitly want to note that this model could work at the level of scientific artifacts (effectively, a digitally enriched form of work that currently only manifests as a paper), but also one step up, at the level of scientific insight that may source from different papers and other intellectual contributions. In scientific insight, we wish to stand on the shoulders of giants, and build upon earlier work. As such, we may source from other insights, similarly to how open-source software may make use of existing other libraries. Furthermore, again looking at functionalities offered by Git, if serious new contributions to an existing repository possibly warrant branching-off into a new strand of independent development, forking functionality allows for this, while again still keeping a living reference to the original repository.

Where in science, the insights we build upon may still be under active research, and there are chances they still may change and update, the same holds for open-source software libraries. This may create a dependency hell, for which software engineering research is actively researching best practices to still make a complex artifact building upon other artifacts as maintainable as possible. We argue that a translation of these best practices will be beneficial in navigating how scientific insight building upon other insights can best be organized and updated, in case all insights dynamically will keep evolving.

As for how to decompose and (re)organize complex code, the software engineering research community has consolidated a rich body of best practices or practices to avoid, consolidated in the form of software engineering methods, design patterns and smells. Equivalents can be formulated for the organization of scientific insight: what sub-experiments or analyses can be modularized or refactored for better reuse? Here, we would like to point out that software engineering methods tend to be taught as advanced-level programming knowledge, and as such may not as actively be part of the skillset of non-computer scientists who took an introductory programming courses---while we believe they are essential in thinking strategically about overall information organization.

\subsection{Intermediate releases with consensus, and organizational safety to find weaknesses, iterate and improve}
When developing a software artifact, pushing code to production, and having formal versioned releases, we naturally agree we have not yet reached The Ultimate and Optimal Final Product---rather, what currently is running may be a Minimum Viable Product that is iterated upon, but that will likely still have many imperfections in need of improvement. In scientific publishing, we may acknowledge this in text, but there is less incentive to demonstrate progressive improvement over subsequent contributions. Furthermore, as we will argue below, it may culturally be unsafe to admit weaknesses and visibly correct them, as this could lead to retractions and consequences on citation track records. In software development, this however is no problem, as changes and releases that can be referenced by others are more clearly separated.

To us, a scientific paper could be seen as a versioned release---a larger, but coherent collection of changes and reviewer consensus that can be frozen and referred to. Similarly, through containerization, we can freeze, save and share the entirety of a computational environment associated to a contribution. If available on a cloud-based platform, this allows for reproducibility, as well as immediate, rapid progress on both the review of material and its reuse and further development, since installation overhead will be reduced. At the same time, these freezes do not signal the end of development, and development can still actively continue.

In our argument to not only organize scientific artifacts as open-source artifacts, but even group them at a higher level of scientific insight, the concept of currently agreed-on consensus can also be taken one level up, similar to how knowledge is established in Wikipedia: for a research problem that many people work on in parallel, a meta-scientific overview of what the collective insight and consensus currently is can be consolidated. Consensus-focused publications aim to condense the overall state of a thread of research, reporting first any consensus, while also indicating ambiguities or research opportunities. One might further conceive `living' consensus-focused articles, i.e. systematic reviews, in a model similar to Wikipedia articles, where the review stays current, as authors continuously update it. This especially will be relevant for topics with increasingly unwieldy numbers of associated publications, where there is a clear benefit to finding a means to condense scientific information; even more so, when the consolidated insights may be looked-at in informing policy (e.g., with regard to climate change or public health).

As software is developed under pressure and with short development timelines, compromises and simplifications will be made. This may lead to technical debt, in which issues needing deeper attention may pile up without being prioritized---up to the point that major and expensive fixes may be needed. Similarly, we would argue that current problems with self-correction in science and questionable insights may be a consequence of intellectual debt, as also suggested in~\cite{zittrainintellectualdebt}. Again, creating cultures in which this is as actively mitigated as possible will be beneficial.

Here, we already mentioned problems with organizational and cultural safety in admitting weakness and implementing corrections in scientific contributions. With software artifacts, we up-front acknowledge that programmers are competent~\cite{DeMilloLS78}, but bugs, errors and issues may still have occurred. Through testing (and ideally, test-driven development) we include safeguards that help reducing the amount of problems that will need fixing---or otherwise will help us signaling and fixing them as early as possible. Still, we will never know whether a program will be fully bug-free. However, this does not prevent us from having justifiable perspectives on when code artifacts can be published and released. We feel that the culture of encouraging and appreciating testing and quality assurance in software engineering may be very inspirational to discussions on fostering Responsible Research Practices and academic integrity, without this being seen as a reputational threat or attack on character.

Finally, the software engineering community has increasingly been acknowledging the social and contextual organizational surroundings of a software artifact, with emerging strands of research studying how team interactions and organizational policies will affect the quality of a software artifact, as well as the efficiency and effectiveness of the process leading to its development. Similarly, these social and organizational insights will be beneficial in efforts to address the culture of scientific research itself (as well as constructive directions for systems changes, if we indeed would decide to move beyond our current ways of sharing insights through static papers).

\section{Conclusion}
In this work, we have outlined how the development of scientific insights parallels the development of software. The shift from traditional publishing to open science involves challenging culture and systems changes. As the software engineering community has noticed in its own open science endeavors, investing in this is a serious, expensive and so far under-rewarded investment~\cite{Mendez2020,artifactsharingse,hermann_artifact_evaluation}. Yet, as we argued, the strong expertise of software engineering experts in acknowledging contributorship on complex larger collaborations, designing for robust maintainability, and developing based on iterative improvements, can more broadly benefit the development of scientific insight subscribing to the Mertonian norms of science---and be beneficial to society at large when complex societal challenges are addressed.

We therefore call upon our software engineering colleagues committed to open science to both think more boldly in how academic incentives can be improved beyond the focus on output, and even look beyond the software engineering research field alone. As for the first, beyond current (commendable) efforts to integrate open science principles in the publication process of software engineering venues, it will be worthwhile to think of what a `Many Labs' equivalent in software engineering may look like. As one thought, may it take aspects of current tool competitions and benchmarks, that also focus on collective understanding, but rather from the start be framed as a joint collaborative and iterative effort?

As for the second, we invite our colleagues to join existing scientific reform movements, help developing and increasing interoperability of current tools, and critically reflect on what software engineering skills can best be taught outside of the own curriculum. As one possible thought experiment, what would a re-framing of the state-of-the-art in climate science as a complex software-like artifact look like? Which insights would need to be decomposed? Who reviews what, and what would a review discussion look like if multiple disciplines get involved? How can we allow for public scrutiny, while not feeding into public distrust?

As one example, we as authors of this article have been actively attending meta-scientific and science improvement events (such as the meetings of the Society for the Improvement of Psychological Science\footnote{\url{https://improvingpsych.org/}}), and have started prototyping the idea of turning scientific publication processes into Git-supported software artifacts~\cite{alexandria}. First prototypical development towards the latter mission was performed in the form of a software development project, which several bachelor students in Computer Science and Engineering at our institute took up as part of their software engineering coursework. The resulting work was presented as a non-archival contribution at a Scientific Progress Seminar~\cite{alexandria_students}. While this was not yet a formal publication, it was an excellent way to get bachelor-level software engineering students interested in research processes, and many of them enthusiastically attended the seminar, that was highly interdisciplinary, also e.g.\ involving epistemological philosophical work. Currently, we are working with our local Open Science community and advertising new student projects to further develop this project.

However, we ourselves are no software engineering researchers, and we are certain our colleagues with deeper expertise in the subject matter can push such developments much further. In doing this, we would argue that software engineering expertise can have even broader societal and scientific impact than it already does today.

\section*{CRediT author statement}
\textbf{Cynthia C.~S.\ Liem}: Conceptualization, Investigation, Methodology, Supervision, Writing – original draft, Writing – review \& editing; \textbf{Andrew M.\ Demetriou}: Conceptualization, Investigation, Resources, Writing – original draft, Writing – review \& editing.

\bibliographystyle{IEEEtran}
\bibliography{references}

\end{document}